\documentclass[aps,pra,superscriptaddress,twocolumn]{revtex4-1}
\usepackage{amsmath,amssymb,graphicx}
\usepackage{subfig}
\usepackage{float}
\usepackage{epstopdf}

\begin{document}


\title{Slow light with electromagnetically induced transparency in cylindrical waveguide}



\author{Agus Muhamad Hatta}
\email[]{ahatta@jazanu.edu.sa}
\affiliation{Department of Physics, Faculty of Science, Jazan University, Jazan, Saudi Arabia}
\affiliation{Department of Engineering Physics, Institut Teknologi Sepuluh Nopember, Surabaya, Indonesia}

\author{Ali A. Kamli}
\affiliation{Department of Physics, Faculty of Science, Jazan University, Jazan, Saudi Arabia}

\author{Ola A. Al-Hagan}
\affiliation{Department of Physics, Faculty of Science, King Khalid University, Saudi Arabia}

\author{Sergey A. Moiseev}
\affiliation{ Kazan Quantum Center, Kazan National Research Technical University,  Kazan, Russia}

\date{\today}

\begin{abstract} Slow light with electromagnetically induced transparency (EIT) in the core of cylindrical waveguide (CW) for an optical fiber system containing three-level atoms is investigated.
The CW modes are treated in the weakly guiding approximation which renders the analysis into manageable form.
The transparency window and permittivity profile of the waveguide due to the strong pump field in the EIT scheme is calculated.
For a specific permittivity profile of the waveguide due to EIT, the propagation constant of the weak signal field and spatial shape of fundamental guided mode are calculated by solving the vector wave equation using the finite difference method.
It is found that the transparency window and slow light field
can be controlled via the CW parameters.
The reduced group velocity of slow light in this configuration is useful for many technological applications such as optical memories, effective control of single photon fields, optical buffer and delay line.

\end{abstract}


\maketitle

\section{Introduction}
Electromagnetically induced transparency (EIT) is a quantum interference phenomenon in which the absorption of a medium can be highly reduced within a frequency band of transparency window \cite{pt-1997}. In EIT system, a three level atomic system is considered where there are two fields, namely a strong control pump field and a weak signal field, interact with the medium. The pump coherent field changes the refractive index of the medium through which the weak signal field propagates and this leads to low absorption of the signal field. In addition, within the transparency window, a dispersion of the medium varies rapidly and it induces significantly reduced group velocity which leads to slow light \cite{n-1999}. Slow light in a medium can be utilized for optical signal processing such as optical buffer, data storage, delay lines \cite{opn-2006}.  Due to the capability to change the susceptibility or refractive index, the EIT scheme has been also explored for several applications such as an all-optical tunable mirror, a wavelength filter, a wavelength monitor, an all-optical switch, a quantum surface plasmon resonance system, and etc. \cite{pier-2008, apa-2012}.

In terms of the medium's field propagation, as an advancement to the bulk medium \cite{pt-1997}, EIT in cavity and waveguides have been investigated \cite{ol-1998, pra-2013}. EIT in cavity can yield very strong pulling of combined cavity-atom resonance frequency towards the EIT resonance frequency \cite{ol-1998}. EIT in planar waveguide has been investigated as \cite{jmse-2011}. In \cite{prl-2008}, EIT medium was proposed at an interface between two materials for an all-optical control of surface polaritons. EIT inside the core of cylindrical waveguide with metal or metamaterial cladding has been considered for possible plasmonic devices applications \cite{pra-2013}. These authors considered slow light for TM surface modes that occur in cylindrical waveguides (with metal or metamaterial cladding), with reduced  loss for the metamaterial cladding.

 In an optical fiber system, the slow light of the fundamental guided mode can be realized using several schemes such as: a birefringent optical fiber and other standard telecom components \cite{prl-2004},  stimulated Brillouin scattering \cite{oe-2005-1}, and Raman assisted fiber optical parametric amplifier \cite{oe-2005-2}. Using EIT based slow light in the optical fibers as well as in the waveguides promises to be very useful for deterministic control of the light fields and as a basic building block for fiber-optic communication  and convenient tool for the light-atoms interactions enhancement. In this paper, we focus on the effect of cylindrical waveguide (CW) on the properties of EIT and the slow light due to spatial inhomogeneous pump coherent field and transverse spatial confinement of the resonant three-level atomic system in the optical fiber. Unlike the work in \cite{pra-2013}, our work here is concerned with EIT and slow light when both the pump and signal fields are in the fundamental guided mode of CW with dielectric core containing resonant three-level atoms and dielectric cladding rather than surface modes and metal or metamaterial claddings. For our analysis we adopt the weakly guiding approximation \cite{kk-2001} where the difference in the refractive indices of core and caldding is assumed very small. This is true in many practical fiber optical applications where the refractive index mismatch is very small. This approximation is useful and reduces the mathematical analysis to more manageable forms. Our focus in this work is on the optimal conditions for the transverse light field confinement in the system under consideration to realize enhanced light-atom interaction that could be easily integrated with conventional optical fiber network. After this introduction we present the model and basic formulation and equations. Results and discussions are given in the following section and the paper concludes with summary of main results.
 
 \begin{figure*}
\centering
\subfloat[]{\label{fig1a}\includegraphics[width=0.45\textwidth]{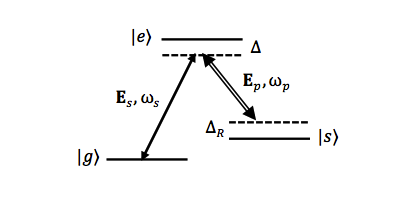}}\qquad
\subfloat[]{\label{fig1b}\includegraphics[width=0.45\textwidth]{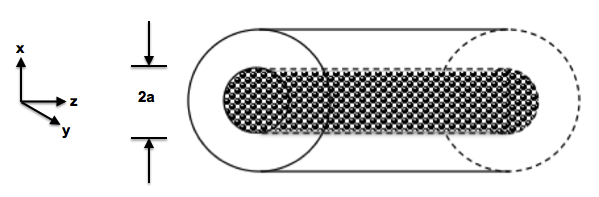}}\qquad
\subfloat[]{\label{fig1c}\includegraphics[width=0.5\textwidth]{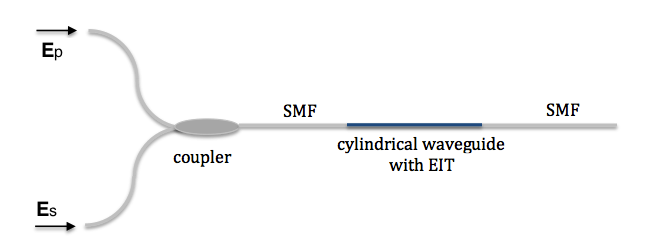}}
\caption{\label{fig1} (a) A schematic of the three-level $\Lambda$ atom, (b) a schematic ofcylindrical  waveguide (CW) where the core embedded with the three level $\Lambda$ atom, (c) Pump and signal fields are injected to the optical fiber system}
\end{figure*}

\section{Formulations and Equations}
A three level atomic system in the $\Lambda$ configuration is considered as shown in Fig. 1(a). A weak signal electric field ${\bf{E}}_\text{s}$ interacts with the atoms on the $|\text{g}\rangle$ $\leftrightarrow$ $|\text{e}\rangle$ transition, and a strong pump coherent electric field ${\bf{E}}_\text{p}$ drives the atomic transition $|\text{s}\rangle$ $\leftrightarrow$ $|\text{e}\rangle$ with Rabi frequency $\Omega_\text{p}$. The three-level $\Lambda$ atom system is embedded in the dielectric core of CW with a surrounding dielectric cladding as depicted in Fig. 1(b). The two fields of pump and signal are injected via an optical fiber system where the fields are coupled into an input singlemode fiber (SMF) as shown in Fig. 1(c). The input SMF is aligned to the CW and again is directed along CW axis to the output SMF to be used for any further signal processing applications in the optical fiber system.

In the optical fiber and CW, the core has radius $a$ and permittivity $\epsilon_1$, and is  surrounded by a dielectric cladding of permittivity $\epsilon_2$. The electric field in the optical fiber and CW is expressed in cylindrical coordinates as \cite{kk-2001}
\begin{align} \label{1}{\widetilde{{\bf{E}}}}(r,\phi,z,t)={\bf{E}}(r,\phi)e^{j(\omega t-\beta z)}
\end{align}
where $\omega$ is an angular frequency of the field and $\beta$ is the propagation constant to be determined by solving a dispersion relation. In this paper, we are concerned with CW fundamental mode. In the weakly guiding approximation where the difference in the refractive indices for the core and cladding is very small the fundamental mode is designated HE$_{11}$ mode or LP$_{01}$  \cite{kk-2001}. The dispersion relation for HE$_{11}$ mode (with the undoped core i.e. in the absence of the three level atoms) is determined by solving the expression  \cite{kk-2001}
\begin{align}\label{2} \frac{J_0(u)}{uJ_1(u)}=\frac{K_0(\text{w})}{\text{w}K_1(\text{w})}\end{align}
where $J_i(u)$, $K_i(\text{w})$ are i$^{\text{th}}$-order of  first kind of Bessel functions and first kind of modified Bessel functions, respectively \cite{kk-2001}. The parameters $u^2$ and $\text{w}^2$ are defined as
\begin{align}\label{3} u^2 =a^2(k_0^2\epsilon_1-\beta^2)\end{align}
\begin{align}\label{4} \text{w}^2 =a^2(\beta-k_0^2\epsilon_2)\end{align}
and related by $u^2+\text{w}^2=V^2$, where $V=k_0a\sqrt{\epsilon_1-\epsilon_2}$ is a normalized frequency, $k_0=\omega/c$ is the free space wavenumber with $c$ is the speed of light in vacuum, and $\omega$ is the field frequency. The electric field is most conveniently expressed in terms of Cartesian coordinates. The field profile of the HE$_{11}$ mode in the transverse component is assumed to be linearly polarised and can be expressed  by \cite{kk-2001}
\begin{align}\label{5}{E_x(r)}&=
\begin{cases}
\frac{E_0}{J_0(u)}J_0\left(\frac{ur}{a}\right),		&\text{ $r\leqslant a$}\\
\frac{E_0}{K_0(\text{w})}K_0\left(\frac{\text{w}r}{a}\right),		&\text{ $r > a$}\\
\end{cases}
\end{align}
where $E_0=\frac{\text{w}}{V}\frac{J_0(u)}{J_1(u)}\left(\frac{2z_0}{\pi a^2 \epsilon_2^{1/2}}\right)^{1/2}$ is an amplitude coefficient and $z_0$ is a vacuum impedance.

\begin{figure}[htb]
\centerline{\includegraphics [height=20cm,angle=0,width=9cm]{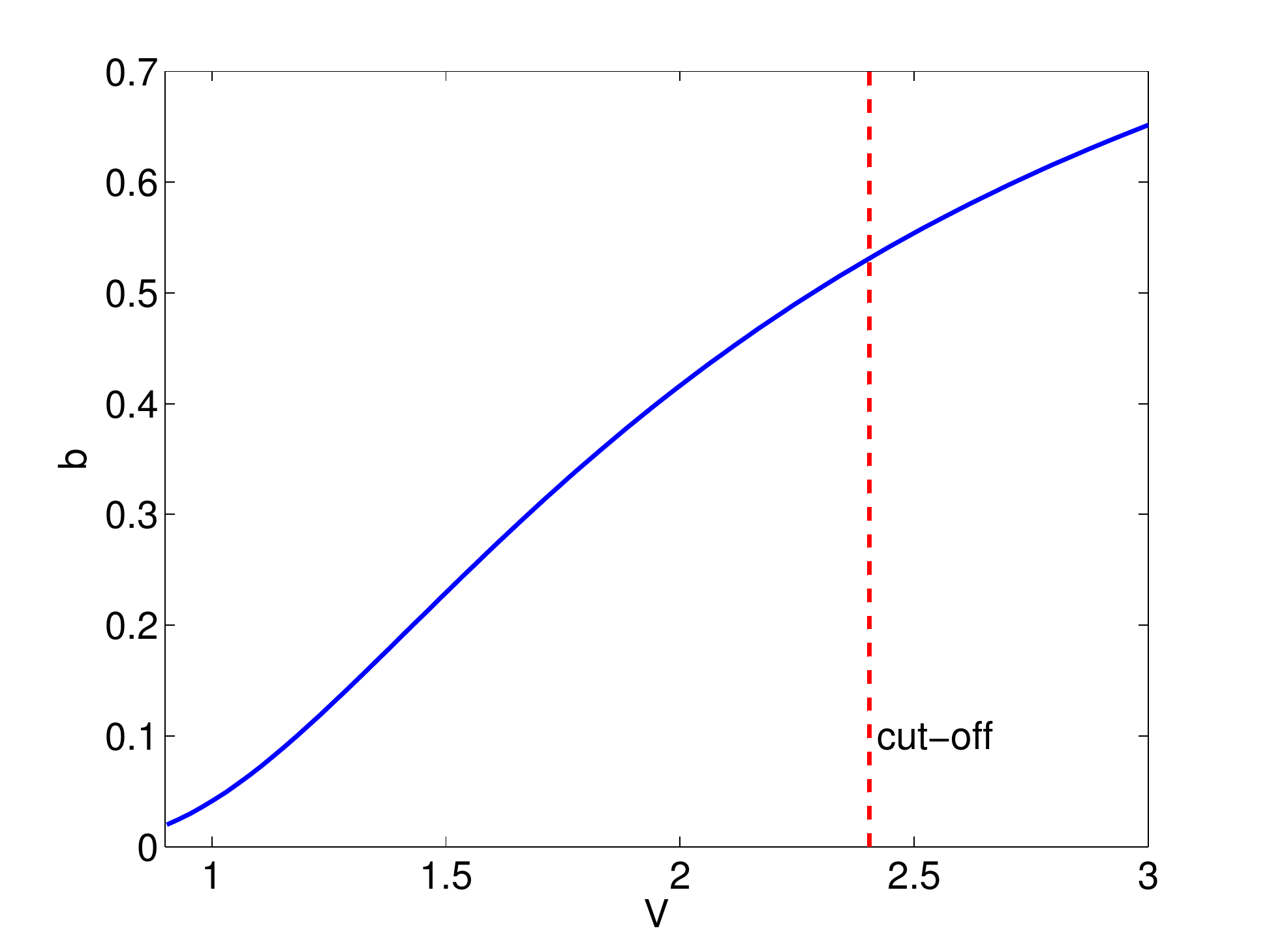}}
\caption {\label{fig2} (Colour online) Dispersion relation of cylindrical or fiber waveguide as a normalized frequency varied with the normalized effective index.} \end{figure}
\begin{figure*}
\centering
\subfloat[]{\label{fig3a}\includegraphics[width=0.45\textwidth]{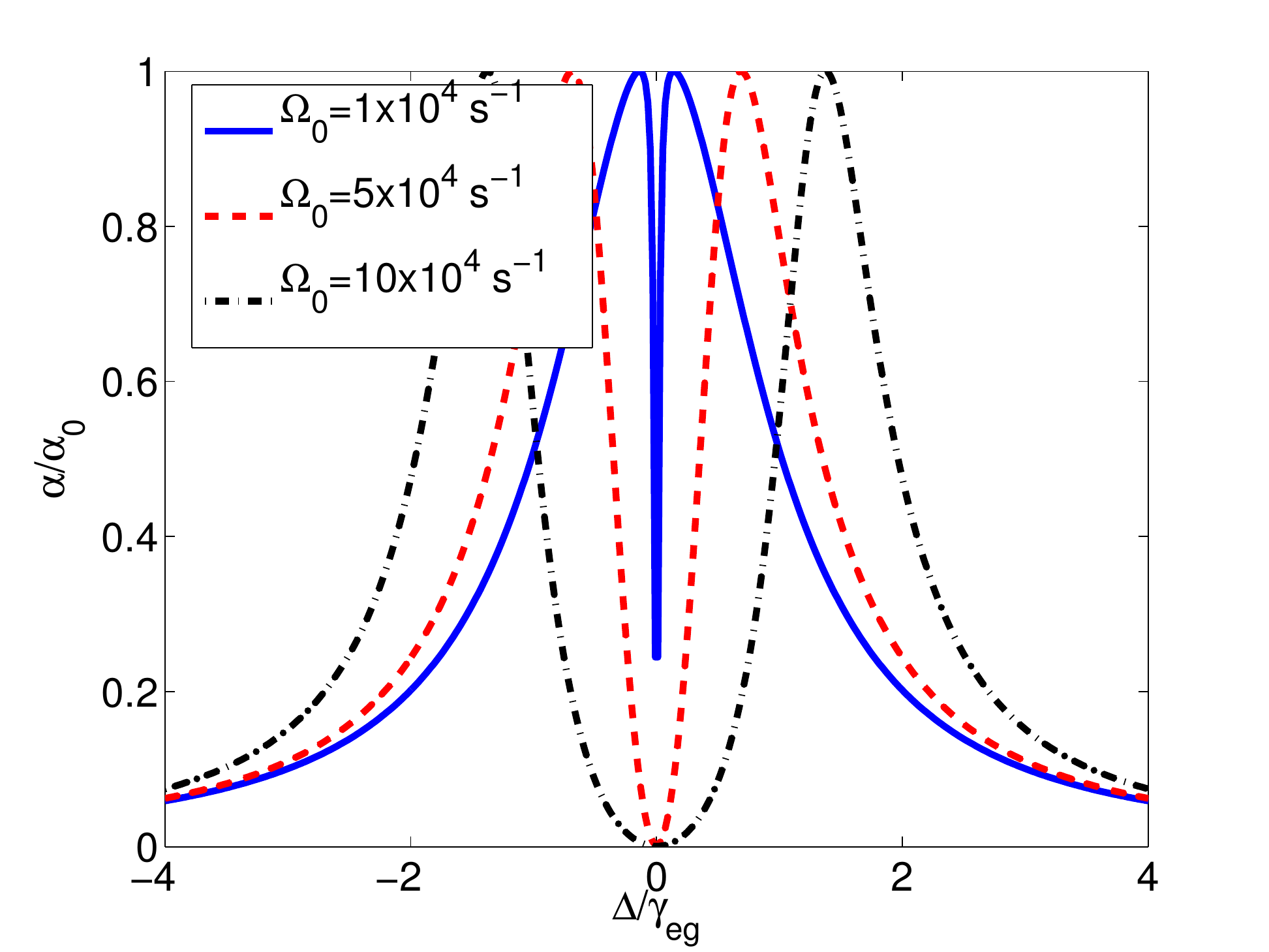}}\qquad
\subfloat[]{\label{fig3b}\includegraphics[width=0.45\textwidth]{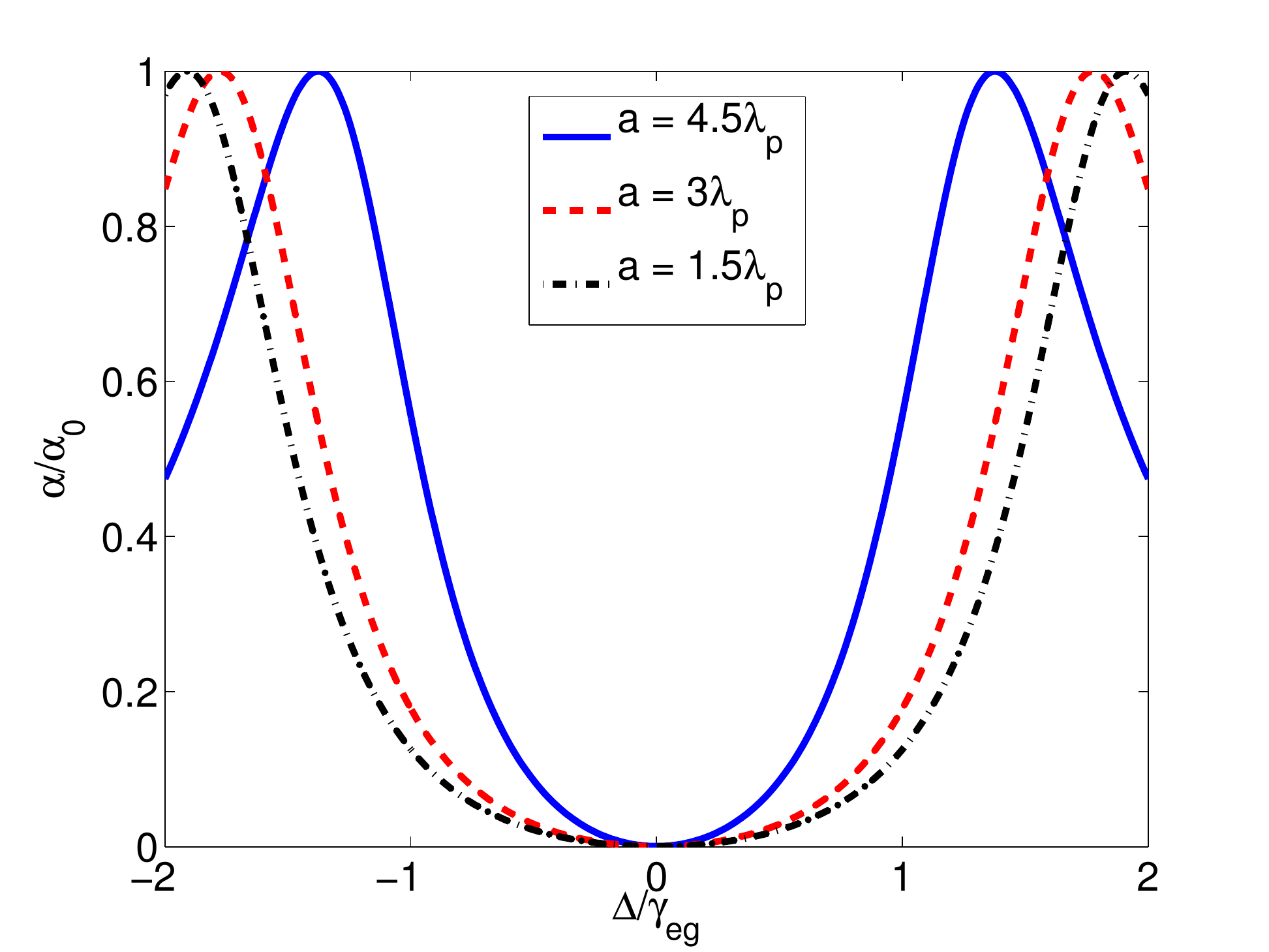}}
\caption{\label{fig3} (Colour online) EIT transparency window for (a) variation of the amplitude of pump field of $\Omega_0$ with a core radius $a$ = 4.5 $\lambda_p$, (b) variation of core radii with a fixed $\Omega_0=10\times10^{4}$ $\text{s}^{-1}$}
\end{figure*}

Equation \eqref{2} is solved numerically for different waveguide structure parameters and frequency. The dispersion relation can be plotted as the normalized frequency $V$ versus a normalized effective index $b=\frac{(\beta/k_o)^2-\epsilon_2}{\epsilon_1-\epsilon_2}$.
Figure 2 shows the dispersion relation for the fundamental guided mode of the cylindrical waveguide in the weakly guiding approximation. It should be noted that for the fundamental mode, the waveguide structure parameters and frequency should satisfy the condition such that the normalized frequency is below a cutoff $J_0(V)=0$ or $V=2.4048$ \cite{kk-2001}. In this paper, the fiber and cylindrical waveguide with undoped core are designed to propagate the fundamental mode only.

So far we have been discussing the undoped weveguide structure. Now, we dope the waveguide core with three level-atoms in the $\Lambda$ configuration which will induce EIT effects. The presence of three-level atoms inside the core will lead to modifications in the core permittivity and is expected to lead to interesting features which we like to explore next. A procedure to model EIT in the waveguide has been presented in refs. \cite{jmse-2011, pra-2013}, and it is adapted in this formulation. The susceptibility of the undoped core is $\chi_0 = \epsilon_1/\epsilon_0 -1$. The presence of the pumped three-level $\Lambda$ atoms under EIT in the core changes the susceptibility into $\chi_0+\chi(\omega,r)$. The susceptibility $\chi(\omega,r)$ can be written as  \cite{sp-2007}
\begin{align}\label{6}\chi(\omega,r) =\frac{2a_0}{n_\text{d}\omega} \frac{i\gamma_{\text{eg}}}{\gamma_{\text{eg}}-i\Delta+|\Omega_\text{p}(r)|^2(\gamma_{\text{sg}}-i\Delta_\text{R})^{-1}},\end{align}
where $n_\text{d}=\sqrt{\epsilon_1 \mu_1}$ is the refractive index of the core with $\mu_1$ the permeability of the core, $\gamma_{ij}$ is the atomic decay rate from level $i$ to level $j$, and
\begin{align}\label{7} a_0 =\frac{3\pi}{n_\text{d}^2\omega^2}\rho_\text{a} \approx \frac{3\pi}{n_\text{d}^2\omega_{\text{eg}}^2}\rho_\text{a} \end{align}
with $\rho_\text{a}$ the number density of three-level atoms, and $\omega_{\text{ij}}$ is the transition frequency for the $|\text{i}\rangle$ $\leftrightarrow$ $|\text{j}\rangle$ transition. The detunings are $\Delta = \omega_{\text{eg}}-\omega_{s}$,  $\Delta_\text{R}  = \Delta -\omega_{\text{es}}+\omega_\text{p}$. The signal field frequency is $\omega_\text{s}$ and pump field has frequency $\omega_\text{p}$ and is set on resonance with the $|\text{s}\rangle$ $\leftrightarrow$ $|\text{e}\rangle$ transition such that $\Delta_\text{R}=\Delta$. The Rabi frequency of the pump is
\begin{align}\label{8} \Omega_\text{p}(r)=\frac{1}{\hbar} {\bf d} \cdot   {\bf{E}}_\text{p}(r), \end{align}
where ${\bf d}$ is the dipole moment of the  $|\text{s}\rangle$ $\leftrightarrow$ $|\text{e}\rangle$ transition, and ${\bf{E}}_\text{p}(r)$ is the pump field given by expression \eqref{5}. For a given pump field in the core, the Rabi frequency of the pump field becomes
\begin{align}\label{9} \Omega_\text{p}(r)=\Omega_0 J_0\left(\frac{u_pr}{a}\right), \end{align} where $\Omega_0=\frac{E_0d}{J_0(u_\text{p})\hbar}$ is the amplitude of the pump field with ${d=\bf d \cdot \hat{r}}$ lies along the field direction, $u_\text{p}^2=a^2\left(k_{\text{es}}^2\epsilon_1-\beta^2\right)$, and $k_{\text{es}}=\omega_{\text{es}}/c$.

The change of core susceptibility due to three level atoms will modify the dispersion and absorption coefficient $(\alpha)$ of the doped core in a manner that depends on the pump field, and can be calculated from the real and imaginary parts of the $\chi(\omega,r)$, respectively. The permittivity of  the waveguide with three-level $\Lambda$ atoms embedded in the core becomes
\begin{align}\label{10}\epsilon_{\text{eff}}\left(\omega, r \right)&=
\begin{cases}
\epsilon_0\left[1+\chi_0+\chi\left(\omega,r\right) \right], 	&\text{ $r<a$}\\
\epsilon_2,	&\text{ $r\geqslant a$.}\\
\end{cases}\end{align}

The weak signal field ${\bf{E}}_\text{s}$ propagates through the waveguide with an appropriate permittivity profile of the cylindrical waveguide due to the EIT. Since the permittivity profile at the core has non-step index profile, the eqs. \eqref{2} and \eqref{5} can not be used to solve the propagation constant of the signal field. Therefore, a vector wave equation should be solved to obtain the propagation constant. It is convenient to write the transverse component of the signal field and the permittivity profile in the Cartesian coordinate as ${\bf{E}}_\text{s}(x,y)$ and $\epsilon_{\text{eff}}(\omega,x,y)$, respectively; and the vector wave equation can be written as
\begin{widetext}\begin{align}\label{11} \bigtriangledown^2{\bf{E}}_\text{s}+\bigtriangledown \left(\frac{1}{\epsilon_{\text{eff}}(\omega,x,y)} \bigtriangledown \left(\epsilon_{\text{eff}}(\omega,x,y)\cdot{\bf{E}}_\text{s} \right) \right) + \epsilon_{\text{eff}}(\omega,x,y) k_0^2 {\bf{E}}_s=\beta^2{\bf{E}}_\text{s}. \end{align}\end{widetext}

A finite difference method (FDM) can be used to solve \eqref{11} and obtain the propagation constant $\beta$ and the corresponding field profile of the signal field \cite{oe-2002}.  With the propagation constant determined from equation (11) for a frequency range, the group velocity can be calculated using
\begin{align}\label{12} v_\text{g}=\left(\frac{\text{d}\beta}{\text{d}\omega}\right)^{-1}. \end{align}

It is clear that the weak signal propagation constant $\beta$ and hence group velocity $v_\text{g}$ depend on the new core susceptibility due the interaction of the three level atoms with signal and pump fields. The wave equation above will be solved numerically using the FDM and results will be presented in the next section.

\begin{figure*}
\centering
\subfloat[]{\label{fig4a}\includegraphics[width=0.45\textwidth]{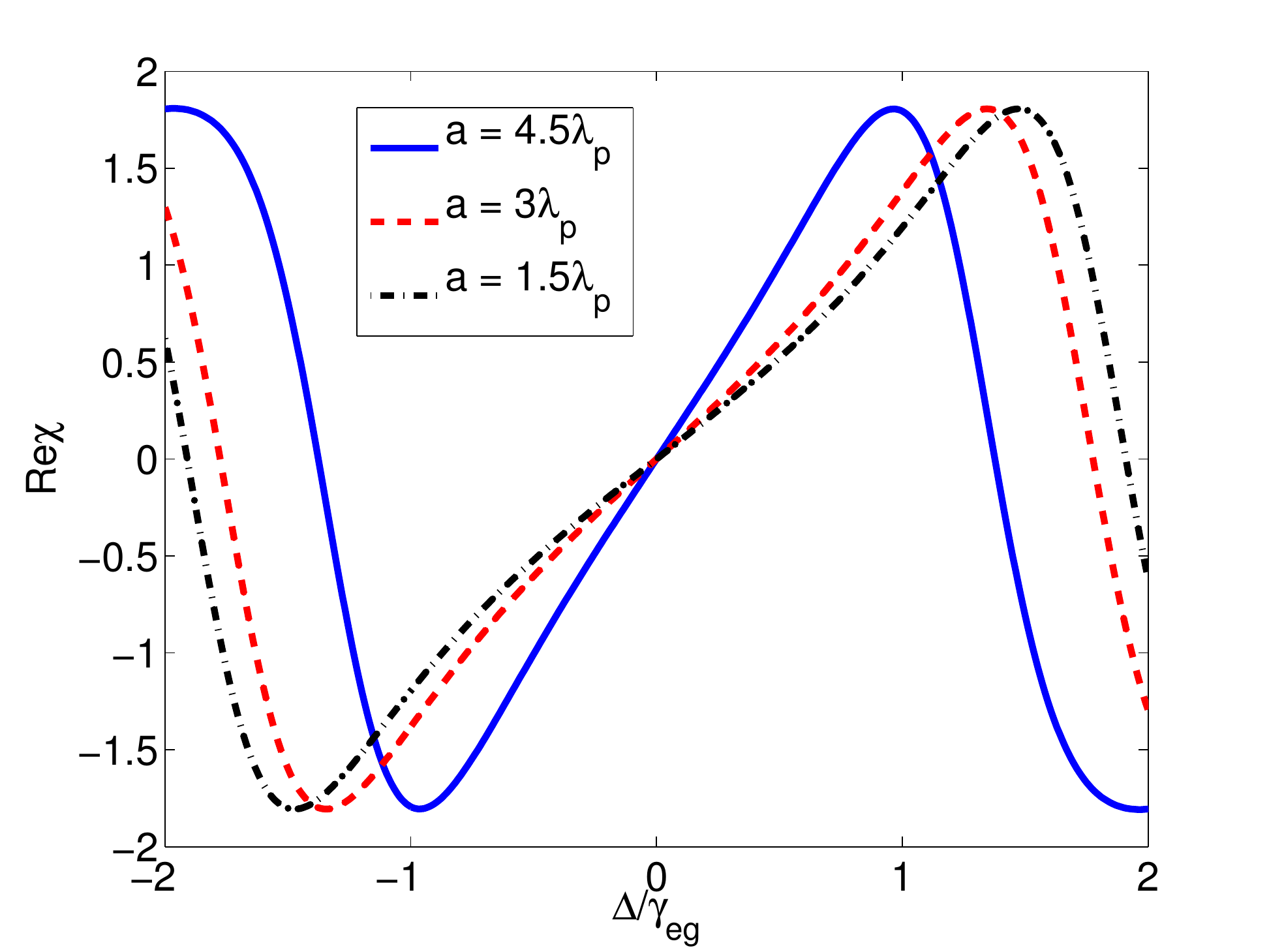}}\qquad
\subfloat[]{\label{fig4b}\includegraphics[width=0.45\textwidth]{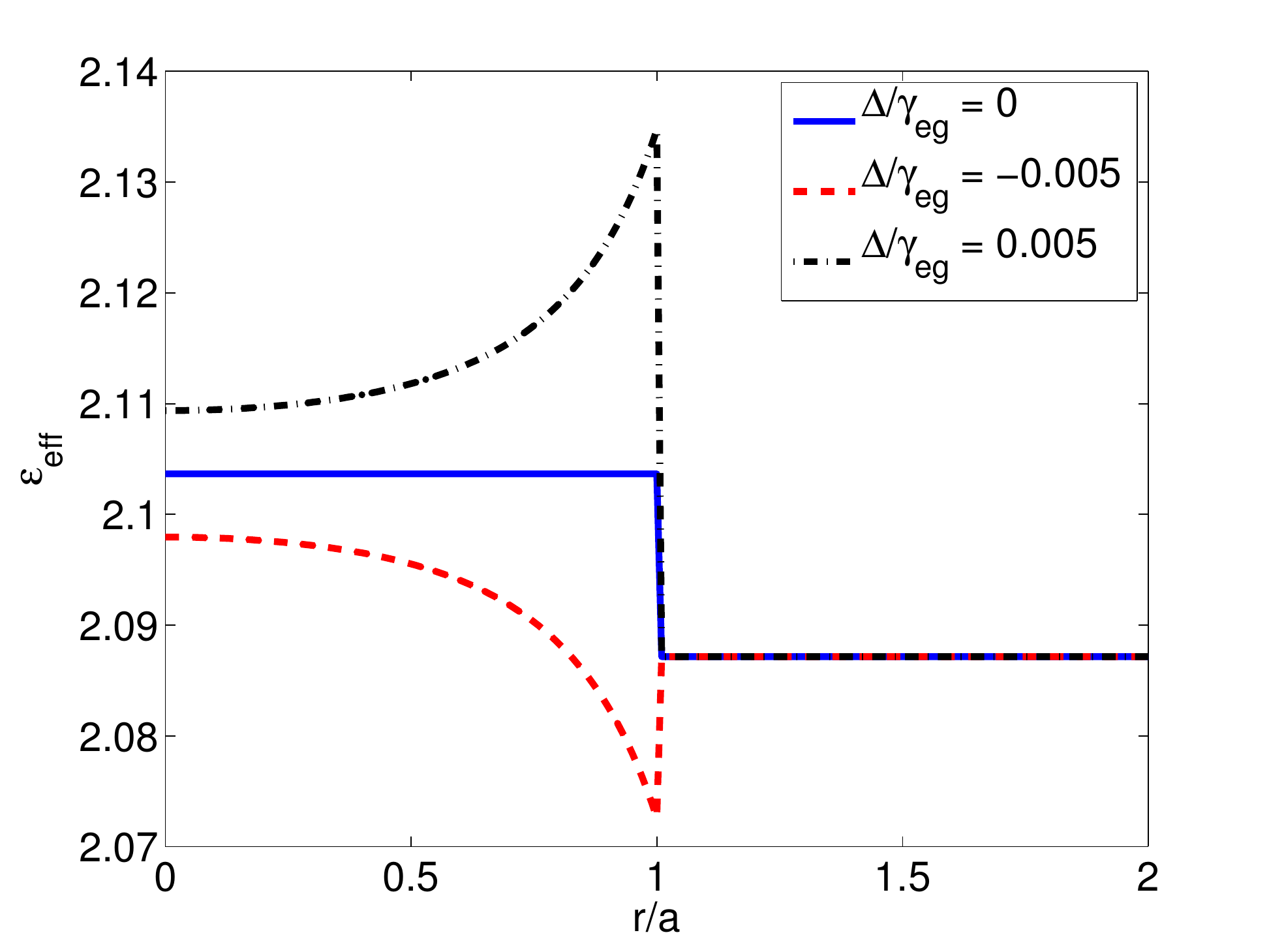}}
\caption{\label{fig4} (Colour online) (a) The dispersion for various core radii at a fixed  $\Omega_0=10\times10^{4}$ $\text{s}^{-1}$, (b) the permittivity profile for $\Delta/\gamma_{eg} = -0.005, 0, 005$, at $\Omega_0 =10\times10^4$ $\text{s}^{-1}$ and the core radius $a = 3\lambda_p$ }
\end{figure*}

\section{Results and Discussions}
For the three-level atoms embedded in the core, we select a material system of Pr$^{3+}$ ions embedded in bulk Y$_2$SiO$_5$ crystal and under cryogenic conditions as in ref \cite{pra-2013}. To calculate the absorption coefficient, dispersion, and permittivity profile, we use some parameters as follows: the core and cladding parameters are $\epsilon_1=2.1037\epsilon_0$, $\epsilon_2=2.0872\epsilon_0$, the core radius $a$ which is comparable to $\lambda_\text{p}=2\pi c/\omega_\text{p}=0.336$ $\mu \text{m}$.
For the three-level atoms embedded in the core, we set the number of density $\rho_\text{a} = 1.26\times10^{15}$ $\text{cm}^{-3}$. The decay rates are $\gamma_{\text{eg}}=1\times10^{5}$ s$^{-1}$, $ \gamma_{\text{sg}}=5\times10^{2}$ s$^{-1}$. The pump and signal frequencies are $\omega_\text{p}=\omega_{\text{es}}$ and  $\omega_\text{s}=\omega_{\text{eg}}$, respectively.

\begin{figure}[htb]
\centerline{\includegraphics [height=20cm,angle=0,width=9cm]{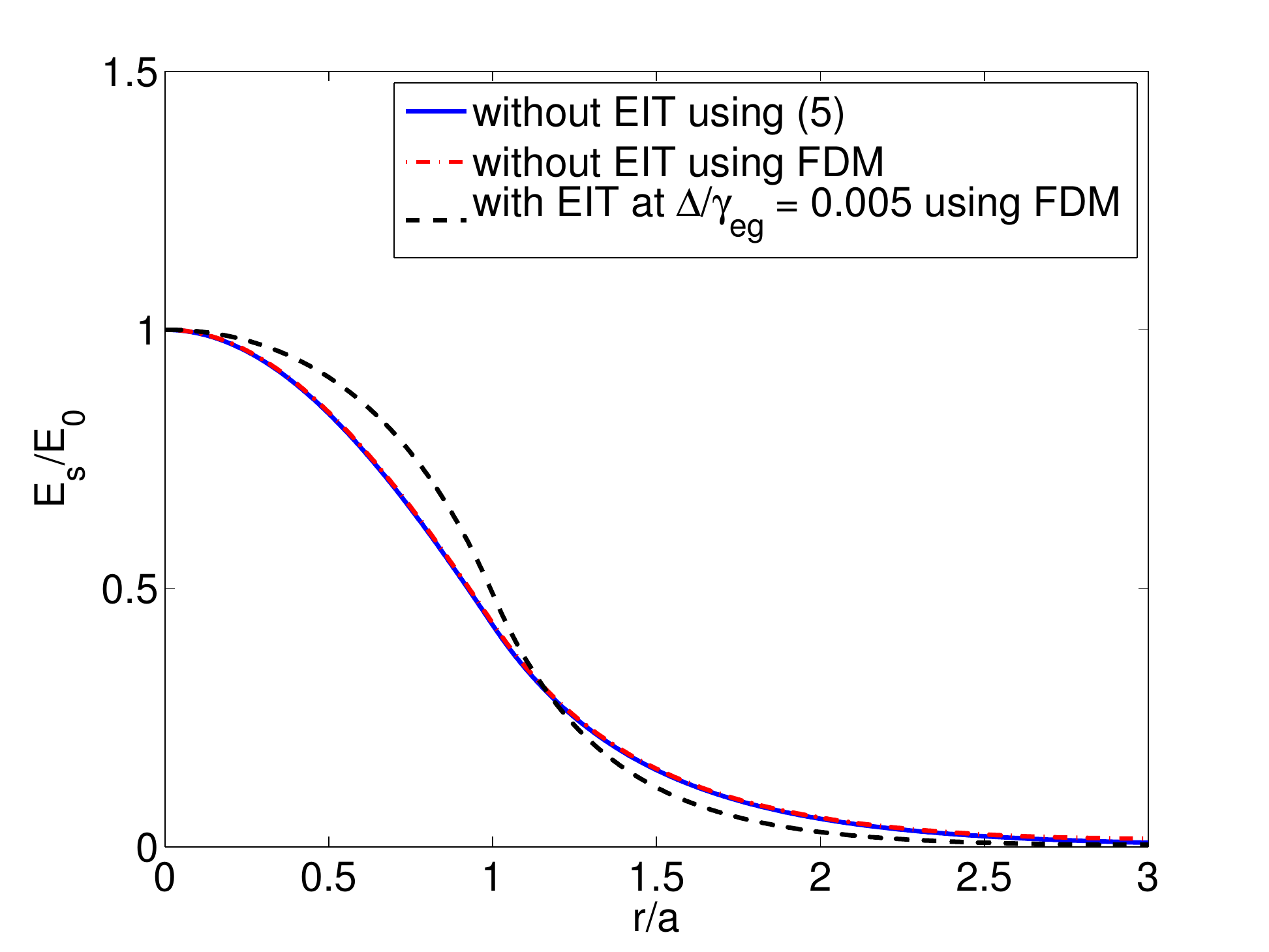}}
\caption {\label{fig5} (Colour online) Normalized signal field profiles without- and with-EIT.} \end{figure} 
\begin{figure*}
\centering
\subfloat[]{\label{fig6a}\includegraphics[width=0.45\textwidth]{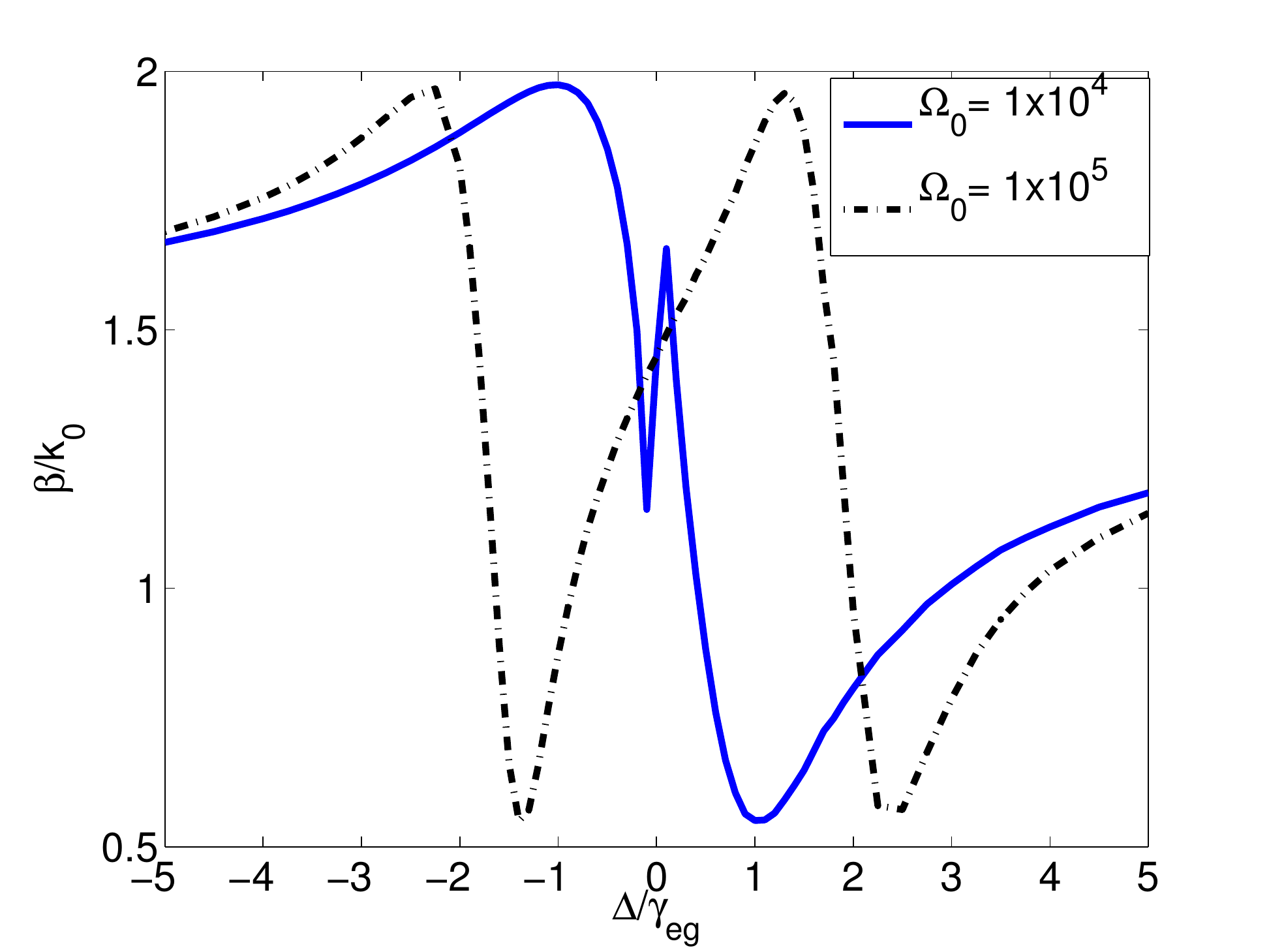}}\qquad
\subfloat[]{\label{fig6b}\includegraphics[width=0.45\textwidth]{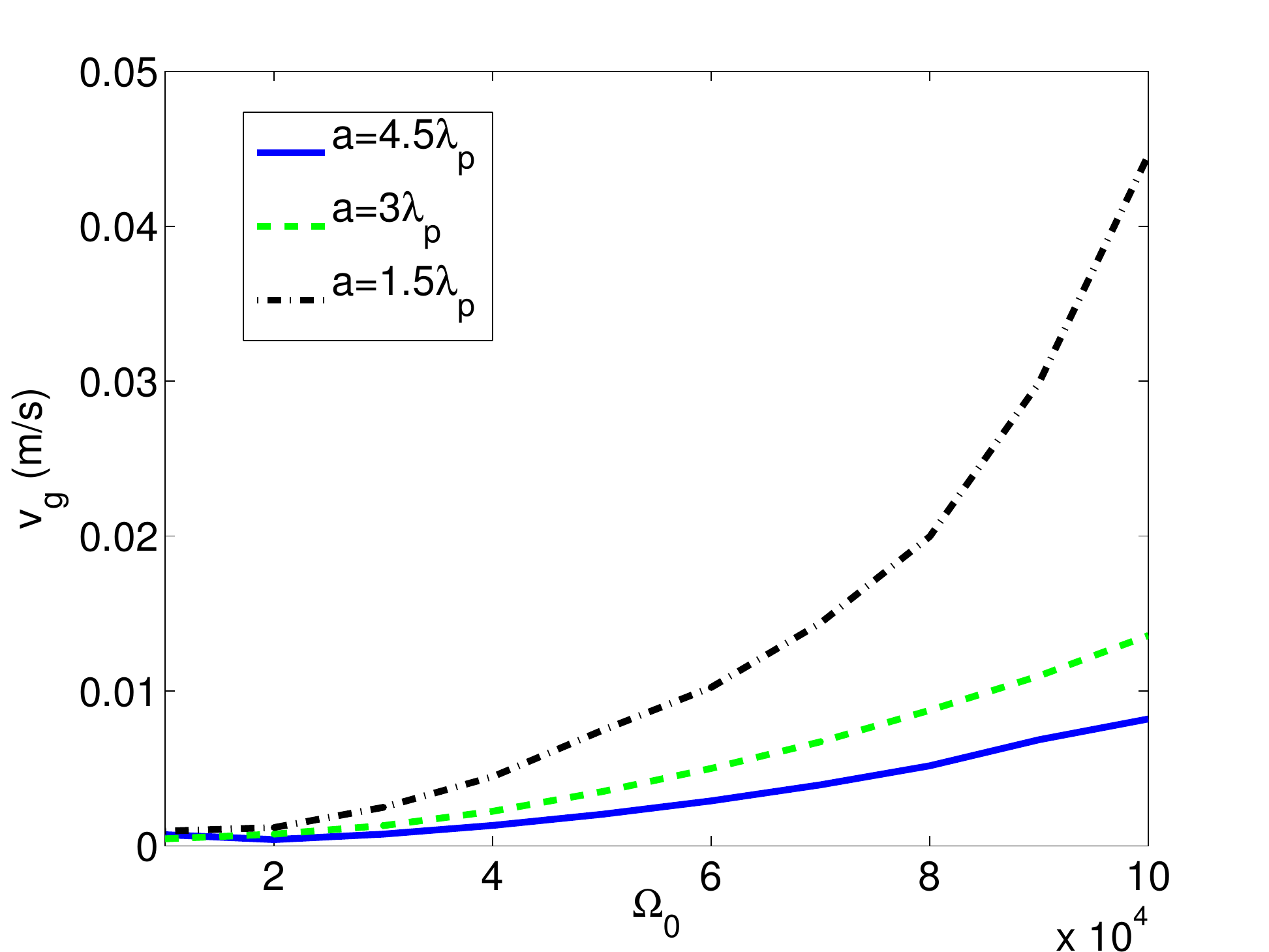}}
\caption{\label{fig6} (Colour online) (a) The dispersion relation of signal field for the core radius $a = 3\lambda_p$, $\Omega_0 = 1\times10^4$ $\text{s}^{-1},$ and $1\times10^5$ $ \text{s}^{-1}$ , (b) the group velocity as a function of $\Omega_0$ for three core radii $a = 1.5\lambda_p, 3\lambda_p,$ and $4.5\lambda_p.$ }
\end{figure*}

For a specific Rabi frequency of the pump field as in equation \eqref{9}, the absorption spectrum $\alpha/\alpha_0$ can be obtained by calculating the imaginary part of \eqref{6}.
Figure 3 shows the EIT transparency window at the centre of the core embedded with the three level $\Lambda$ atom as a function of the normalized detuning. In Fig. 3(a), the interaction with the pump field results in a splitting of the absorption spectrum into two peaks, and the line centre of the medium becomes transparent to the resonant signal field. For a fixed core radius $a=4.5\lambda_\text{p}$, the transparency window can be broadened by increasing the amplitude of pump field $\Omega_0$ as well known in EIT. The transparency window can also be increased by reducing the core size as depicted in Fig. 3(b), for the core radii $a=4.5\lambda_\text{p}, 3\lambda_\text{p},$ and $1.5\lambda_\text{p},$ and a fixed $\Omega_0=10\times10^4$ $\text{s}^{-1}$.

Figure 4(a) shows the dispersion given by the real part of $\chi(\omega,r)$ for several core radii at fixed $\Omega_0$ and the same parameters as in Fig. 3(b). One can see the dispersion yields a steep and approximately linear slope at the transparency window near  the resonance frequency $\Delta = 0.$  The dispersion depends on the core radius while increasing the core radius increases the slope gradient. Figure 4(b) shows the permittivity profile for several normalized detunings $\Delta/\gamma_{\text{eg}}=-0.005,0,0.005$ at $\Omega_0=10\times10^4$ $\text{s}^{-1}$ and the core radius $a=3\lambda_\text{p}$. For the normalized detuning $\Delta/\gamma_{\text{eg}}=0$, the core permittivity has a step function profile , whereas for non-zero detunings $\Delta/\gamma_{\text{eg}}\ne 0$, the core permittivity shows a graded profile.

When the signal field propagates through a given permittivity profile as in Fig. 4(b), the field profile and propagation constant of the signal field can be evaluated by solving equation \eqref{11} using the FDM. Figure 5 shows the normalized signal field profiles without and with EIT. One can see that for the fundamental mode, the signal field profile without EIT, calculated using the weakly guiding approximation as given by equation \eqref{5} and the field obtained from more exact FDM are in a good agreement. For the EIT condition at $\Delta/\gamma_{\text{eg}} = 0.005$, the permittivity in the core becomes higher and the graded profile is as shown previously in Fig. 4(b), and therefore the signal field is more confined inside the core compared to the condition without-EIT.

Figure 6(a) shows the calculated propagation constant within a range of normalized detuning for the core radius $a = 3\lambda_\text{p}$, $\Omega_0 = 1\times10^4$ $\text{s}^{-1},$ and $1\times10^5$ $ \text{s}^{-1}$. The propagation constant varies rapidly near the resonance, with steeper slope for lower Rabi frequency $\Omega_0$. Once the dispersion profile obtained, the group velocity can be computed using equation \eqref{12} and it is depicted in Fig. 6(b). The group velocity is displayed as a function of $\Omega_0$ for three core radii $a = 1.5\lambda_\text{p}, 3\lambda_\text{p},$ and $4.5\lambda_\text{p}$. The group velocity can be controlled and slowed down by reducing the Rabi frequency $\Omega_0$. However, for a smaller $\Omega_0$, the absorption coefficient is higher at the transparency window as shown in Fig. 3(a). The group velocity also depends on the core radius, it decreases as the core radius gets larger.

Thus, the slow light of signal field  in cylindrical waveguide can be controlled by adjusting the amplitude of pump field of $\Omega_0$ which is determined from the solutions of CW fundamental guided mode. It also depends on the core radius of the cylindrical waveguide. The light can be almost stopped for a choice of lower value of Rabi frequency $\Omega_0$, and it leads to potential applications such as optical memories, optical buffer, and optical delay line. Meanwhile, the core radius has also important role to control the slow light and the bandwidth of the transparency window.

\section{Conclusions}
We have investigated the EIT inside the core of CW in the optical fiber system in the weakly guiding approximation where the core and cladding indices difference is very small. This approximation turns out to practical and convenient tools for the anaylisis of EIT inside complex CW system. The results agree very well with more elaborate finite difference calculations for the lowest fundamental mode considered here. Higher order modes can also be treated but are not considered in this work. For EIT inside CW it is found that the transparency window can be broadened by increasing the amplitude of pump field of fundamental mode and decreasing the core radius of the CW containing three level atomic system. For a given permittivity profile due to the EIT, the propagation constant of the signal field can be calculated by solving the vector wave equation using the finite difference method. The group velocity of the signal field can be reduced by increasing the core radius of the CW, and by decreasing the Rabi frequency parameter as EIT demands. This study is useful for slow light development in the optical fiber system. The analyzed scheme can be easily integrated with conventional optical fiber systems at the conditions of enhanced light-atom interaction due to transverse light field confinement. It can be also useful for realization of enhancement of light atom interaction which is essential for optical quantum repeaters and other quantum optical communications.

S.A.M. acknowledges  the Russian scientific fund through Grant No. 14-12-01333 for financial support.


\end{document}